\documentstyle[12pt]{article}
\let\useblackboard=\iftrue
\iftrue
\def\nbaselineskip{15pt}
\fi
\useblackboard
\font\blackboard=msbm10 scaled \magstep1
\font\blackboards=msbm7
\font\blackboardss=msbm5
\newfam\black
\textfont\black=\blackboard
\scriptfont\black=\blackboards
\scriptscriptfont\black=\blackboardss
\def\Bbb#1{{\fam\black\relax#1}}
\else
\def\Bbb{\bf}
\fi

\def\sect#1{\subsection{#1}\setcounter{equation}{0}}
\catcode`@=11
\def\@cite#1#2{\if@tempswa [#1]\else$^{\scriptscriptstyle
\mbox{\rm\scriptsize#1}}$\fi}
\newcommand{\eqn}{\begin{eqnarray}}
\newcommand{\enq}{\end{eqnarray}}
\newcommand{\eqa}{\begin{array}}
\newcommand{\ena}{\end{array}}
\newcommand{\eq}{\begin{equation}}
\newcommand{\en}{\end{equation}}

\newcommand{\no}{\nonumber}

\def\comments#1{}

\def\1N{$1\over N$}
\def\CC{\Bbb{C}}
\def\CP{$\Bbb{P}^n$}

\def\CCC{\Bbb{C}}
\def\sCP{$\Bbb C{\rm P}^n$}

\def\G24{$\Bbb{G}(2,4)$}

\def\ci{\Bbb{C}}

\def\CS{\Bbb{P}}

\def\ZZ{\Bbb{Z}}
\def\del{\partial}

\def\Tr{{\rm Tr\ }}

\def\ket#1{|#1\rangle}
\def\vev#1{\langle{#1}\rangle}

\begin{document}
\setlength{\unitlength}{0.25cm}
\begin{titlepage}
\hfill{\vbox{\hbox{{\sc OUTP-95-34P}}
\hbox{\sc CERN-TH/95-338}
\hbox{ hep-th/9512149}}}
\vspace{1.5cm}

\begin{center}
{\LARGE
Supersymmetric $\sigma$-Models on Toric Varieties:\\[0.2in]
A Test Case.}\\[0.4in]

{
Mich\`ele Bourdeau
\footnote{{\tt bourdeau@surya11.cern.ch}}}\\
CERN-TH\\
1211 Geneva 23\\[1.5cm]
\end{center}

\vfil
\begin{center}
{\sc Abstract}
\end{center}

\begin{quotation}

In this letter we  study supersymmetric $\sigma$-models on  toric
varieties. These manifolds are generalizations of \CP\
manifolds.
We examine here  $\sigma$-models, viewed as  gauged linear $\sigma$-models,
 on one of the simplest such manifold, the blow-up
of $\CS^2_{(2,1,1)}$,  and determine their properties
using the techniques of topological-antitopological fusion.
We find  that the model contains solitons which
become massless at the singular point of the theory
 where a gauge symmetry remains unbroken.

\end{quotation}
\vfill
CERN-TH/95-338\hfill\\
December 1995 \hfill

\end{titlepage}
\setlength{\baselineskip}{\nbaselineskip}
\newpage

\sect{Introduction}

\indent We aim in this paper to study $N=2$ $\sigma$-models on a particular
class of toric varieties, weighted projective spaces,
which are generalizations of ordinary projective spaces.
These spaces have been explored extensively by string theorists, in relation
to string compactifications on Calabi-Yau manifolds.
{}From a string theory point of view, one is typically interested in
Calabi-Yau manifolds that can be realized as
hypersurfaces defined by polynomials in weighted projective
spaces.\cite{cls,yau,sw,kky,kl,mgvwll} These conformal models form a
large class of consistent string vacua.
More recently, considerations of mirror symmetry have led physicists
and mathematicians to  study the ambient space of the Calabi-Yau manifold,
or the full toric variety.\cite{batyrev, mopl}

These nonlinear $\sigma$-models with K\"ahler target space
can be obtained as the low-energy limit of certain two dimensional
$N=2$ supersymmetric models with abelian gauge symmetry,
or gauged linear $\sigma$-models (GLSM), as shown by Witten.\cite{wit1}
These gauged linear $\sigma$-models can be twisted to give topological models,
which in turn can  be used to calculate instanton expansions for correlation
functions in topological nonlinear $\sigma$-models.\cite{mopl}

Weighted projective spaces contain orbifold singularities. We can however
replace each singular point of the singular locus   by a $\CS^1\sim S^2$
to give a smooth K\"ahler manifold.\cite{co}
These singularities may cause
some of the correlation functions to diverge at these points
 because the infinite instanton sums needed to calculate the correlators
can contain singularities.

Recently there has been a great deal of interest concerning the nature
of the appearance of physical singularities at points in moduli
spaces and this question has been examined in $N=2$ susy gauge theories and
 type ${\rm I\!I}$ string theory compactifications on various manifolds.
The explanation is
argued to be that the apparent singularity is due to
 nonperturbative massive states becoming massless at
these singular points.\cite{seiwit, stro}

Here we will explore the physics of  field theories
which contain similar types of singularities, supersymmetric
$\sigma$-models on toric varieties in two spacetime dimensions.
One interesting question is whether we can make any
predictions regarding the behavior of the theory near the
singular points.

For this, we look at $\sigma$-models on one of the simplest such
toric variety, the space obtained by resolving the singularity of
 the weighted projective space $\CS^{2}_{(2,1,1)}$ or ${\CS^2}/{\ZZ_2}$.
We view these models as gauged linear $\sigma$-models and
explore their properties  using the powerful techniques of
topological-antitopological fusion developed by Cecotti
and Vafa.\cite{cv1,cv2,cv3} These methods  allow to study
various characteristics of the model along the whole renormalization
group flow.
The parameters defining the flow are
the two couplings $\beta$ and $\alpha$ of the GLSM Lagrangian.
These couplings are related to the two homology cycles corresponding to the
$\CS^2$ and to the blowing up of its singular point.
The correlation functions of the model depend on these two parameters.
The one singularity of the model appears as a pole in the correlation
function at $\beta=1/4$.
We find, by flowing to the infra-red, that the non-linear
$\sigma$-model contains Bogolmonyi
solitons whose masses are determined to be proportional to $(1-4\beta)^{1/2}$.
This shows that, as we tune the coupling $\beta$ towards its singular
value of $1/4$, the solitons of the model become massless.
At this point, a continuous gauge symmetry of the GLSM is restored and gives
rise to flat directions. Thus
we have another example of a field theory where the apparent physical
singularity can be explained by a gauge symmetry enhancement with
the appearance of massless  states. Here, due to the relative simplicity
of the model, we can explicitly see what happens.

The organization of the paper is as follows. We  start by
reviewing the geometrical data
needed to define the toric variety and its classical cohomology.
 In section three, we review the  Lagrangian
description of non-linear supersymmetric $\sigma$-models on toric
varieties,  obtain the quantum cohomology and  discuss some of the
physical properties of the model.
In section four we determine the topological-antitopological fusion
 equations which we will then  solve  in particular cases.
The first case  is a solution of the model when we
set $\beta$ to zero (section five).
This case might be seen as an interesting Landau-Ginsburg model on its own,
but we mainly give the solution as an example of a completely solvable case.
Next, in section six, we set out to determine the behavior of the model
near the singularity. This turns out to be  possible because of a
 simplication of the equations when we take $\beta$ to be real.
We obtain, through the
asymptotic behavior of the equations, the masses of the solitons
 and find their dependence on the parameters.
Finally we summarize our results and their consequences.

\sect{The toric data for the blow-up of $\CS^2_{(2,1,1)}$}

This section is rather technical and is a review of material found in
various papers (see \cite[x,y,z,w,q,r]{Fulton,cls,mopl,agm,cox,batyrev}).

Toric varieties can be defined in terms of simple combinatorial data.
This geometrical data encodes information about various aspects of the
space, such as its  cohomology.

\noindent The toric variety $V$ corresponding to a complex weighted
projective space is described as
\eq
\CS^{(n-1)}(k_1,k_2,\dots,k_n)= \CCC ^{n+1}- \{0\}\,/\,\CCC ^*
\en
with $\CCC^*$ acting by $\,\lambda \dot
(z_1,\dots,z_n)=(\lambda^{k_1}z_1,\dots, \lambda^{k_n}z_n),\,$
for all nonzero $\lambda$, where $(z_1,\dots, z_n)$ are local holomorphic
coordinates on $V$.
These spaces have orbifold singularities, due to the identifications
$ (z_1,\dots,z_n)\simeq(\lambda^{k_1}z_1,\dots, \lambda^{k_n}z_n),$
except for the case when all weights are unity,  which corresponds to
ordinary projective space \CP. In fact,
\eq
\CS ^{n-1}_{(k_1,\dots,k_n)}= {{\CS ^{n-1}}\over{\ZZ
_{k_1}\times\dots\times\ZZ_{k_n}}}
\en
as can be seen by setting $z_j=(\zeta)^{k_j}$ such that
$\,(\zeta_1,\dots,\zeta_n)\simeq \lambda(\zeta_1,\dots,\zeta_n),\,$ and
identifying $\,\zeta_j\simeq e^{2\pi i k_j}\zeta_j.\,$
These identifications lead to singular sets.

  A more general definition of a $n$-dimensional  toric
variety  is as the quotient space\cite{mopl}
\eq
V=(Y-F_{\Delta})/T_{\Delta}
\en
where $Y=\CCC^n,\, T_{\Delta}\sim \CCC^{*(n-d)}$  acts diagonally on
the coordinates of $Y$ by
\eq
g_a(\lambda ):x_i \rightarrow \lambda ^{Q_i^a} x_i\qquad
a=1,\dots ,(n-d),\quad i=1,\dots ,d,
\en
and  $F_{\Delta}$ is a subset of $Y - $\CC$^{*d}$ which is a union of
certain intersections of coordinate hyperplanes.
The combinatorial data $\Delta$ defining $V$ determines the
intersection of coordinate hyperplanes and the integers $Q^i_a$ which
specify the representation of the gauge group.\cite{Fulton,mopl,agm,cox}

The weighted projective space $\CS^2_{(2,1,1)}$ is identified with
${\CS^2}/{\ZZ_2}$. Indeed, consider the three
complex homogeneous coordinates $(z_1,z_2,z_3)$ describing the space.
One has $\,(z_1,z_2,z_3)\simeq
(\lambda^2z_1,\lambda z_2,\lambda z_3)\,$
which, with $\lambda = -1$,  becomes $\,(z_1,z_2,z_3)
\simeq (z_1,- z_2,- z_3).\,\,$ Take now a neighborhood of the point
$(1,0,0);\,$ there is a $\ZZ_2$ identification on the space and the
action fixes $(1,0,0)$. The singular set consists of the point
$(1,0,0)$.
In order to obtain a smooth projective space, this point is blown up
to a $\CS^1$. This process of blowing up introduces new $(1,1)$ forms
which contribute (with the K\"ahler form) to the cohomology of the
smooth toric variety.

The toric data\cite{Fulton, agm, mopl} can be described in
terms of a `fan' which is a
collection of two-dimensional cones $\sigma_i$. Each cone is spanned
by two one-dimensional cones (or simplices), and each one-dimensional
cone can be identified with a complex variable $z_i$.
The one-dimensional simplices for the weighted projective space
$\CS^2_{(2,1,1)}$ are given by\cite{Fulton} the vectors
 $\{\vec{v_1}=(1,0)\equiv z_1,\, \vec{v_2}= (0,1)\equiv z_2,\, \vec{v_3}=
(-2,-1)\equiv z_3\}.$
The area defined by the triangles formed by two such vectors and the
origin are not equal. Removing the singularity involves  making these
areas equal and this is done by adding the vector $\,\vec{v_4} =
(-1,0)\equiv z_4.$ The vector is called the `exceptional divisor' in the
language of mathematicians and is the average
$(\vec{v_1}+\vec{v_2})/2$.
The set $F$ contains points which are removed so that there are no
fixed points. Here
\eq
F=\{z_1=z_4=0\}\cup \{z_2=z_3=0\}.
\en
($F$ is defined by taking the union of all sets obtained by setting to
zero all coordinates corresponding to vectors of a primitive
collection. A primitive collection consists of vectors not generating
a single cone.)

We now need to mod out by the action of $T=(\CS^*)^2$.
Let $D \in \ZZ^n$ be the sublattice of vectors
$d=(d_1,\dots,d_n)$ such that $\sum_i d_iv_i=0$. Choosing a basis
$\{Q^1,\dots,Q^{n-d}\}$ for $D$ gives the $T$ action.
In our case,  $Q^1 = (1,0,0,1)$ and $Q^2=(0,1,1,-2).$
Thus
\eqn
g_1(\lambda)&\rightarrow &(\lambda z_1, z_2, z_3, \lambda z_4)\\
g_2(\lambda)&\rightarrow &( z_1, \lambda z_2, \lambda z_3,
\lambda^{-2}z_4)
\enq
We can now obtain the cohomology of the manifold by proceeding
 as explained in \cite[x]{mopl}.  The cohomology $H^*(V)$ is generated
 by $H^2(V)$ under the intersection product.
The group $H^2(V)$ is generated by classes
$\xi_i, \quad i=1,\dots, n,\,$ dual to the divisors $\{x_i=0\}$, subject to
linear relations. The dimension of $H^2(V)$ is $n-d$. A basis $\eta_a$
of $H^2(V)$ is such that
\eq
\xi_i = \sum_{a=1}^{n-d}Q_i^a\eta_a.
\en
In our example,
$\xi_1=\eta_1,\,\,\xi_2=\eta_2,\,\,\xi_3=\eta_2\,$and
$ \,\xi_4=\eta_1-2\eta_2.$

\noindent The nonlinear relations in the ring $H^*(V)$ are determined as
follows.
For each irreducible component of $F$, described as $\{x_a=0\, |\, a\in
A\}$ for some set $A\in \{1,\dots,n\}$, there is a relation
$\prod_{a\in A}\xi_a=0.$
These produce the following classical ring relations
\eq\label{cl}
\xi_2\xi_3=\eta_2^2 = 0,\qquad \xi_1\xi_4 = (\eta_1-2\eta_2)\eta_1=0.
\en
The correlations functions are also determined by the
toric data. Take a collection of $d$ distinct coordinate hyperplanes
$\{x_i=0\},\dots , \{x_{i_d}=0\}$ which do intersect on $V$, then
$\vev{\xi_{i_1}\dots \xi_{i_d}}_V=1$, if $V$ is smooth.
This implies $\vev{\eta_1\eta_2}=1,\,\vev{\eta_1^2}=2.$

\sect{The nonlinear $\sigma$-model and quantum cohomology}

\indent Recall that the \sCP\ and  Grassmannian $\sigma$-models
 both have descriptions as gauged $N=2$ models\cite{cv2, bd, bou}
\eq\label{lgg}
   {\cal L} = \int d^4\theta \left [ \sum_{i,a} \bar{S}_{ia}
    e^{-V} S_{ia} + \alpha \,\Tr V\right ]
\en
with $a$ a `flavour' SU$(N)$ index and $i$ a `gauge' U$(k)$ index
($a=1\,$ and $\,\alpha \,\Tr V = {{A}\over{2\pi}}V$ for the \sCP model).
The $S_i$ are chiral superfields, and V is a real vector (or matrix)
superfield.

 Integrating out the superfields $S_i$ in (\ref{lgg}) for the \sCP\ models,
one obtains an effective action  which, by gauge invariance, contains
only the field-strength superfields $X$ and $\bar{X}$ since $V$ is
not gauge invariant:
\eqn
S_{\rm eff}={N\over{2\pi}}\!\int\!\!d^2x\left\{
\int \!\! d^2\theta W(X)\!+\!\!
\int\!\! d^2\theta \bar{W}(\bar{X})
\!+\!\!\int\!\! d^4\theta[Z(X,\bar{X},\Delta,\bar{\Delta})]\right\}
\enq
with
\eq
W(X)={1\over{2\pi}} X(\log X^N-N+A(\mu)-i\theta),
\en
$A$ is a renormalized coupling, $\theta$  the
instanton angle,
and
$\,  X=D_L\bar{D}_R V,\,\,\bar{X}=D_R\bar{D}_L V.$

\noindent This action has the form of a Landau-Ginsburg model.
A $N=2$ supersymmetric theory in two dimensions admits a
Landau-Ginsburg description if it has a superspace Lagrangian
such that
\eq
{\cal L}=\int d^4\theta \sum_i \phi_i\bar{\phi}_i +\int d^2\theta
W(\phi_i)+h.c.\qquad
\en
where $\phi_i$, $\bar{\phi_i}$ are  chiral and antichiral
 superfields and the superpotential $W$  is an
analytic function of the complex superfields.
The ground states of the theory are  $dW(\phi)=0$.
The chiral ring is the ring of polynomials generated by the
$\phi_i$
modulo the relations
$dW(\phi)/d\phi_i = D \bar D \phi_i \sim 0$.
(For a review, see \cite[x,y]{lvw,warner}.)

\noindent For the \sCP\ model,  the chiral ring is thus the powers
of $X$ mod $dW=0$, or $X^N = e^{ -A+i\theta} \equiv \beta.$

The Grassmannian $\sigma$-models admit a LG description as well,\cite{cv2,wit2}
 with $W$ given by
\eq\label{lgw}
 W_f(\lambda_1,\lambda_2,\dots,\lambda_k)={1\over {2\pi}}
\sum_{j=1}^k\lambda_j(\log \lambda_j^N-N+A(\mu) -i\theta ).\no\\
\en
The gauge-invariant fields are now polynomials in the eigenvalues
$\lambda_m$ of the field-strengths $\lambda$ and are generated by the
elementary symmetric functions
\eq\label{var}
X_i(\lambda )\equiv\sum_{1\leq l_1
< l_2 < \dots l_i\leq k}\lambda_{l_1}\lambda_{l_2}
\dots \lambda_{l_i}\quad (i=1,\dots, k)
\en
The ring relations are
$\lambda_j^N=$ const. and the quantum cohomology  of the
Grassmannian $\sigma$-models are generated by
the elementary symmetric functions $X_i$'s.

Similarly, for the nonlinear $\sigma$-model with target space the toric variety
$V$, there exists a manifestly $N=2$ supersymmetric gauged linear
$\sigma$-model
with target space $Y$ and gauge group $G=U(1)^{(n-d)}\,$
with $G_{\ci}=T$.\cite{wit1,mopl}

\noindent The Lagrangian is
\eq
{\cal L}=\int d^4\theta\left [ \sum_i \bar{S}_{i}
    e^{(2\sum_{a=1}^{n-d}Q_i^aV_a)} S_{i} -\sum_{a=1}^{n-d}r_aV_a\right ]
\en
where the $n$ chiral matter multiplets $S_i$ with charge $Q_i^a$ under $G$
are coupled to the $n-d$ abelian gauge superfields $V_a$.

\noindent Integrating out the chiral superfields, one is left with
the superpotential
\eq\label{W}
W(\Sigma) = {1\over{2\sqrt{2}}}\sum_{a=1}^{n-d}\Sigma_a\left
(i{\tau_a} -{1\over{2\pi}}\sum_{i=1}^nQ_i^a
\log(\sqrt{2}\sum_{b=1}^{n-d}Q_i^b\Sigma_b/\Lambda)\right)
\en
where $\Sigma_a={1\over {\sqrt{2}}}{\bar{D}_+}D_-V_a$ are the (twisted
chiral) gauge-invariant field strengths associated with the gauge fields
$V_a$ and have component expansions
\eq
\Sigma = \sigma -i\sqrt{2}(\theta^+\bar{\lambda}_++
{\bar{\theta}}^-{\lambda}_-)
+\sqrt{2}{\theta}^+\bar{\theta}^-(D-f)+\dots\no
\en
\noindent As we will see, the gauge-invariant field strengths
will again generate the cohomology
of the model, with relations determined by $dW(\Sigma)=0.$

\noindent The coupling $\tau_a\equiv ir_a+{\theta_a\over{2\pi}}$,
where $r_a$ is a renormalized coupling and $\theta_a$ is the instanton angle.

As stressed in \cite[x]{mopl}, the model reduces in the low energy
limit to the nonlinear
$\sigma$-model with target space $V$ when $r_a$ lie within
a certain cone ${\cal K}_c$ of $V$. This follows from finding
the space of classical
ground states of the theory by setting the potential for the bosonic
fields to zero. Doing so induces a relation on the $r_a$ through the
auxiliary fields $D_a$
\eq
U=\sum_{a=1}^{n-d}{{(D_a)^2}\over{2e^2}} + 2\sum_{a,b=1}^{n-d}\bar{\sigma}_a
\sigma_b\sum_{i=1}^nQ_i^aQ_i^b|\phi_i|^2,\no
\en
where $D_a=-e^2(\sum_{i=1}^nQ_i^a|\phi|^2-r_a)\equiv 0\,$ is the condition
for vanishing energy.

\noindent In our case, the condition for the $r_a$ to lie in ${\cal K}_c$
are  $r_1\geq 0, \,r_2+2r_1\geq 0$.
 The smooth phase, corresponding to $r_a$ in the K\"ahler cone
${\cal K}_V$ is determined\cite{mopl} by the conditions $r_1>0,\,r_2>0$.
It appears classically that the space of vacua reduces to a point for
$r_a\equiv 0$ and that supersymmetry is spontaneously broken for negative
$r_a$ as the energy can no longer vanish. However, quantum mechanically,
one finds  a smooth
continuation to negative $r_a$ with unbroken
supersymmetry.\cite{wit1,wit2,mopl}
The interaction (\ref{W}) and the constraints on the chiral fields derived
from it are thus valid for
 values of $r_a$ outside of ${\cal K}_c$,
 when we analytically continue to other regions in parameter
space, i.e. here for $r_a$ negative. In fact, the formalism of
the topological-antitopological equations
(which we describe in the next section)
does not distinguish between the sign of $r_a$ and
allows naturally to go beyond zero radii and resolve the singularities of
classical geometries.\cite{cv3}

The quantum cohomology of the toric variety is obtained by setting
$dW(\Sigma_a)=0$. This produces the  constraints
\eq
\prod_{i=1}^n\left (\sum_{b=1}^{n-d}Q_i^b\Sigma_b\right )^{Q_i^a} =
e^{2\pi i \tau_a}\equiv q_a, \quad  a=1,\dots,n-d,
\en
and the classical ring relations (\ref{cl}) for $\CS^2_{(211)}$ are changed to
\eq
\Sigma_1(\Sigma_1-2\Sigma_2)= \alpha,\qquad \Sigma_2^2 = \beta
(\Sigma_1 - 2 \Sigma_2)^2,
\en
in the quantum cohomology ring,
with the deformation parameters $q_1\equiv \alpha $,  $\,q_2\equiv \beta$
 functions of the GLSM couplings $\tau_1\equiv ir_1+{\theta_1\over{2\pi}},
\,\tau_2\equiv ir_2+{\theta_2\over{2\pi}}$. The parameters $r_1,r_2$
represent the areas of the two homology cycles corresponding respectively
to one on $\CS^2$ and the exceptional divisor.
The correlation functions become, using the relations in the ring,
\eq\label{correl}
\vev{\Sigma_1\Sigma_2}=1,\qquad \vev{\Sigma_1^2}=2,\qquad
\vev{\Sigma_2^2}={{-2\beta}\over{1-4\beta}}.
\en
We notice here a singularity of the model at $\beta = 1/4.$
As explained  in \cite[x,y]{wit1,mopl}, such a singularity arises because
the instanton sums contributing to the correlator become infinite with possible
singularities if $\,\sum_iQ_i^a=0\,$ (here for $a=2$).
This happens  because there are then solutions to
$D_a=0$ which leave a continuous subgroup of $G$ (here $g_2$) unbroken and
give rise to flat directions ($r_2=0$) where the space of susy ground-states
is non-compact and the theory singular. Quantum corrections will have
the effect of shifting $\tau^a\,$ to
 $\,\tau_{eff}^a=\tau_a\,+\,{i\over{2\pi}}\sum_iQ_i^a\ln Q_i^a\,\,$
and therefore in our  case, these will lead  to a singularity at
$\,\tau_2={i\over\pi}\ln 2,\,\,\theta_2=0,\,$ or $\,\beta=1/4$.

\noindent Other toric varieties of the form $\CS^{n-1}_{(2,\dots,2,1,1)}$
(with $[n-2]$ 2's) will
also have this single singularity and the following results can be assumed
to generalize to these other models.

We are now in a position to study the model using the techniques of
topological-antitopological fusion.

\sect{The $tt^*$ equations for the model}

The $tt^*$ equations describe the way in which a certain hermitian  metric
$g_{i\bar{j}}$ changes
along the renormalization group flow. The topological-antitopological
metric $g_{i\bar{j}}\equiv\vev{\bar{j}|i}$ is just the inner
product on the supersymmetric ground states  $\ket{i}$ and
$\ket{\bar{j}}$ of the theory. These Ramond ground states are in
one-to-one correspondence with the chiral superfields.
The metric can be thought of as a generalization of the Zamolodchikov metric
away from the conformal point. This metric and a new index derived from it are
helpful for  understanding various properties of the model, like the scale and
coupling dependence and the soliton spectrum
(see \cite[x,y]{cv1,bd} for a review).

\noindent As a basis for the chiral ring, which is generated by
$\{\Sigma_1,\Sigma_2\},$ we take
\eq
{\cal R} = \{1,\Sigma_1, \Sigma_2, \Sigma_1\Sigma_2\}.
\en
The topological metric or two-point function $\eta_{ij}$ is related to
$g_{i\bar{j}}$ by the `reality constraints'
\eq
\eta^{-1}g(\eta^{-1}g)^*=1.
\en
The metric is determined by the
following differential equations\cite{cv1}
\eq\label{dif}
  \bar{\partial}_{\bar j}(g\partial_ig^{-1})
  =[C_i,gC^{\dagger}_{\bar j}g^{-1}]
\en
\eq
\del_iC_j -\del_jC_i + [g(\del_ig^{-1}),C_j]-[g(\del_jg^{-1}),C_i]=0.
\en
The $C_i$ represent the action on the chiral ring of
the operators corresponding to a perturbation by the couplings.
In the present case, we have  two couplings $\alpha$ and $\beta$, and
the matrices $C_{\alpha}$ and $C_{\beta}$ are (with $a=-2\beta/(1-4\beta))$
\[ C_\alpha =-{1\over\alpha}\left(\eqa{cccccc}
0 & 1 & 0 & 0 \\
\alpha & 0 & 0 & 2 \\
0 & 0 & 0 & 1 \\
0 & 2\alpha\beta & -\alpha 2\beta/a & 0
\ena \right),\qquad
C_\beta =-{1\over\beta}\left(\eqa{cccccc}
0 & 0 & 1 & 0 \\
0 & 0 & 0 & 1 \\
-\alpha a/2 & 0 & 0 & a\\
0 & \alpha\beta & -2\alpha\beta & 0 \\
\ena \right). \]

\[{\rm \!\!\!\!\!\!\!\!\!\!\!The\,\,two\!-\!point\,\,function\,\,
is,\,\,in\,\,view\,\,of\,\,
(\ref{correl})}
\qquad\qquad\eta =\left(\eqa{cccccc}
0 & 0 & 0 & 1 \\
0 & 2 & 1 & 0 \\
0 & 1 & a & 0 \\
1 & 0 & 0 & 0
\ena \right). \]
$\Sigma_1$ and $\Sigma_2$ having same (complex) dimension, we expect
off-diagonal elements in the metric (with $g_{2\bar{1}}=g_{1\bar{2}}^*$)
\[ g = \left(\eqa{cccccc}
g_{0\bar{0}} & 0 & 0 & 0 \\
0 & g_{1\bar{1}} & g_{1\bar{2}}  & 0 \\
0 & g_{2\bar{1}} & g_{2\bar{2}} & 0 \\
0 & 0 & 0 & g_{3\bar{3}}
\ena \right) \]
The diagonal entries $g_{1\bar{1}}$ will be a function only of
$(|\alpha|,|\beta|)$ since total chiral charge is zero, and chiral
charge non-conservation is proportional to instanton number.
The reality constraints imply
\eqn\label{real1}
&&2(1-2a^*)=-a^*g_{1\bar{1}}^2+2g_{1\bar{2}}g_{1\bar{1}}-2g_{1\bar{2}}^2\\
&&(1-2a^*)=-a^*g_{1\bar{1}}g_{2\bar{1}}+g_{1\bar{1}}g_{2\bar{2}}
+|g_{1\bar{2}}|^2-2g_{1\bar{2}}g_{2\bar{2}}\no\\
&&a(1-2a^*)=-a^*g_{2\bar{1}}^2+2g_{2\bar{1}}g_{2\bar{2}}-2g_{2\bar{2}}^2\no\\
&&1=g_{0\bar{0}}g_{3\bar{3}}\no
\enq
The $tt^*$ equations and reality constraints will in general be simplest
in the so-called `flat basis,'  the basis where the two-point functions are
independent of the couplings $\tau_a$.

\noindent We find the flat basis to be ${\cal R}=\{1, \Sigma_1/ \sqrt{2},
(\Sigma_1-2\Sigma_2)/\sqrt{2}\delta, \Sigma_1\Sigma_2\}$
with $\delta = \sqrt{2a-1}$.

\noindent $\eta $ becomes
 \[\eta = \left(\eqa{cccccc}
0 & 0 & 0 & 1 \\
0 & 1 & 0 & 0 \\
0 & 0 & 1 & 0 \\
1 & 0 & 0 & 0
\ena \right) \]
and the reality constraints simplify to
\eqn\label{real2}
&&g_{1\bar{1}}^2 + g_{1\bar{2}}^2 =1\no\\
&&g_{2\bar{1}}^2 + g_{2\bar{2}}^2 =1\no\\
&&g_{2\bar{1}}g_{1\bar{1}}=-g_{1\bar{2}}g_{2\bar{2}}
\enq
These imply $g_{1\bar{1}}^2=g_{2\bar{2}}^2$ (or $g_{2\bar{2}}=-(+)
g_{1\bar{1}}$ if  $\,g_{1\bar{2}}$ is  real (imaginary))

\noindent Calling $\overline{g_{i\bar{j}}}$ this new basis, we can go
 from one basis to the other
\eqn\label{ch}
\overline{g_{1\bar{1}}}&=&{1\over 2}g_{1\bar{1}},\quad
\overline{g_{0\bar{0}}}=g_{0\bar{0}}\no\\
\overline{g_{1\bar{2}}}&=&{1\over{2\delta^*}}(g_{1\bar{1}}-2g_{1\bar{2}})\no\\
\overline{g_{2\bar{2}}}&=&{1\over{2|\delta|^2}}(g_{1\bar{1}}-2(g_{1\bar{2}}
+g_{2\bar{1}})+4g_{2\bar{2}})
\enq
Making the appropriate changes in the $C_i$ matrices for the new basis,
we can then easily obtain the $tt^*$ equations (\ref{dif}) in the flat basis.

\sect{The solution of the model for $\beta =0$}

The $tt^*$ equations determine solutions for the metric $g_{i\bar{j}}$ from
which one can understand many properties of the model. For example, the
 asymptotic behavior  of the metric in the infra-red
limit gives the masses of the solitons of  the model.\cite{cv3}
The $tt^*$ equations are in general very
complicated and few analytical solutions are known.\cite{cv1,bd,bou}
We find here two
interesting special cases where the equations simplify
and  the model is  solvable.
The first case  is when we set  $\beta \equiv  0.$
This case might be more than just interesting as a boundary condition for the
 general solution of the metric. It might also correspond  to some kind
of Landau-Ginsburg theory.
\noindent Now, the reality constraints (\ref{real1}) can be solved
completely (since $a\equiv 0$) and they imply (if $g_{2\bar{2}}\neq 0$)
\eq\label{c1}
g_{1\bar{1}}=g_{2\bar{2}}+g_{2\bar{2}}^{-1},\qquad
g_{1\bar{2}}=g_{2\bar{2}}.
\en
The $tt^*$ equations reduce to
\eqn\label{tt1}
-\del_{\bar{\alpha}}\del_{\alpha}\ln
g_{0\bar{0}}&=&{1\over{|\alpha|^2}}g_{0\bar{0}}^{-1}\left [
 g_{2\bar{2}}^{-1}+g_{2\bar{2}}\right ] - g_{0\bar{0}} g_{2\bar{2}}\no\\
-\del_{\bar{\alpha}}\del_{\alpha}\ln
g_{2\bar{2}}&=&{1\over{|\alpha|^2}}g_{0\bar{0}}^{-1}\left [g_{2\bar{2}}^{-1}
- g_{2\bar{2}}\right ] - g_{0\bar{0}} g_{2\bar{2}}
\enq
But this system of equations is familiar. Calling $\,q_0=\ln
g_{0\bar{0}},\,\, q_2=\ln g_{2\bar{2}}$, \,\,$q_{02}=q_0 + q_2 = \ln
\,g_{0\bar{0}}\,g_{2\bar{2}}\,\,$ and $\,\,\overline{q_{02}}=q_2 - q_0 = \ln
g_{2\bar{2}}/g_{0\bar{0}},\,\,$
we have
\eqn\label{qo2}
\del_{\bar{\alpha}}\del_{\alpha}\overline{q_{02}}&=&
                 {2\over{|\alpha|^2}}e^{\overline{q_{02}}}\no\\
\del_{\bar{\alpha}}\del_{\alpha}q_{02}&=&
           2\left [e^{q_{02}}- {e^{-q_{02}}\over{|\alpha|^2}}\right ]
\enq
The first of (\ref{qo2}) reduces to the Liouville equation by the
transformation\\
$\ln g_{2\bar{2}}/g_{0\bar{0}}=2\overline{q_{02}}' + \ln |\alpha|^2$:
\eq
\del_{\bar{\alpha}}\del_{\alpha}\overline{q_{02}}'=
                e^{\overline{2q_{02}}'}
\en
whose general  solution is\cite{sacgm}
\eq\label{liouv}
\overline{q_{02}}'=\ln\left[2\left|{{dg}\over{d\alpha}}
\right|{{1\over{(1-gg^*)}}}\right]
\en
where $g(\alpha)$ is an analytic function of $\alpha$.
If we do not want any angular dependance in the solution so that
the metric depends only on $|\alpha|$, we need
$g=\alpha^m$, m real. However, this solution is singular along the
curve $gg^*=1$ and the real nonsingular solution is thus for $m\ne 0$. Then
\eq
g_{2\bar{2}}/g_{0\bar{0}}={{4m^4|\alpha|^{2m}}\over{(1+|\alpha|^{2m})^2}}.
\en
The second of (\ref{qo2}) can also be put in a recognizable
form. Setting $z=4\sqrt {\alpha}$ and
$\quad \ln g_{0\bar{0}}\,g_{2\bar{2}}= q_{02}' -\ln |\alpha|$, we get the
Sinh-Gordon equation
\eq\label{sg}
\del_{\bar{z}}\del_{z}{q_{02}'}= \sinh q_{02}'.
\en
The asymptotic limits of this equation are known.
We  look for solutions which are real, regular and non-zero as
$z\rightarrow 0$.
Following \cite[x]{cv1}, the solutions to the Sinh-Gordon equations
are classified by their asymptotic behavior as $|z|\rightarrow 0\,$ (or
$\,r_1\rightarrow \infty) ,$
\eqn
q_{02}'(|z|)&\simeq & r\log |z| + s + {\cal O}(|z|^{2-|r|})
\qquad {\rm for}\,\, |r|<2\\
q_{02}'(|z|)&\simeq & \pm 2\log |z|\pm 2\log\left[-\left (\log{|z|\over
2}+\gamma\right )\right]+{\cal O}(|z|^4\log^2|z|) \qquad
{\rm for} \,r=\pm 2,\no
\enq
where $r$ and $s$ are
related by $e^{s/2}={1\over {2^r}}{{\Gamma({1\over 2}-{r\over 4})}\over
{\Gamma({1\over 2}+{r\over 4})}},$ and $\gamma$ is Euler's constant.

\noindent In order for the metric to have no poles as a function of $\alpha$,
we need $r=-2$ (and $m=1$).
So
\eq
g_{0\bar{0}}g_{2\bar{2}}=\left[4|\alpha|
          \left(-\ln 2|\alpha|^{1/2}-\gamma\right)\right]^{-2}
\en
and
\eqn
g_{2\bar{2}}^2&=&{1\over 4}\left[(1+|\alpha|^2)\left(-\ln 2|\alpha|^{1/2}
-\gamma\right)\right]^{-2}\no\\
g_{0\bar{0}}^2&=&(1+|\alpha|^2)^2\left[8|\alpha|^2\left(-\ln 2|\alpha|^{1/2}
-\gamma\right)\right]^{-2}.\no\\
\enq
\noindent For $z\rightarrow \infty \,(r_1\rightarrow -\infty )$, the
general solution to (\ref{sg}) is
\eqn
q_{02}'(\alpha)&\sim & -{2\over{\sqrt{\pi}}}\sin (\pi
r/4){1\over{\sqrt{z}}}\exp[-2z]+\dots\no\\
 &\sim&{2\over{\sqrt{\pi}}}{1\over{\sqrt{z}}}\exp[-2z]\no
\enq
Since the leading asymptotic behavior of the metric is determined by the
one-soliton contributions, we have fundamental   solitons
of mass\cite{cfiv,cv3} $m=2|z|$. We get
\eqn
g_{2\bar{2}}^2&=&{{4|\alpha|}\over {(1+|\alpha|^2)^2}}\left[1+\sqrt{2\over\pi}
|\alpha|^{-1/4}\exp{-4|\alpha|^{1/2}}\right]\no\\
g_{0\bar{0}}^2&=&{{|\alpha|}\over {4}}\left[1+\sqrt{2\over\pi}
|\alpha|^{-1/4}\exp{-4|\alpha|^{1/2}}\right].\no
\enq
which suggest that the model contains solitons of mass $4|\alpha|^{1/2}$.

\sect{The behavior of the model near the singularity}

For the general case, the $tt^*$ equations look quite complicated to
solve. As we presently see, we now describe  a  case
where we  can obtain a result  which will allow us to
predict the general behavior of the
theory. This happens when we take the metric to be diagonal in the flat
basis $(\overline{g_{1\bar{2}}}=0).$ In this case $\beta$ is real and
the equations are tractable. Since the one singular point of the theory
occurs at $\beta=1/4$ (or $\tau_2={i\over \pi}\ln 2,\,\theta = 0$),
we can study  what happens in the vicinity of the singularity.

\noindent Recall that if we take $\overline{g_{1\bar{2}}}=0$,
we can have either
$\overline{g_{1\bar{1}}}=\overline{g_{2\bar{2}}}=1$ or
$\overline{g_{1\bar{1}}}=-\overline{g_{2\bar{2}}}=1$ (see \ref{real2}).
If we look at the full $tt^*$ equations in the flat metric, we notice that
the condition for a solution to exist when $\overline{g_{1\bar{2}}}=0\,\, $
is $\,\,\delta - \delta^*=0\,\, $ if $\,\overline{g_{1\bar{1}}}
=\overline{g_{2\bar{2}}}=1\,\, $
and $\,\,\delta + \delta^*=0\,$ when $\,\overline{g_{1\bar{1}}}
=-\overline{g_{2\bar{2}}}=1$.
  In either case, for $\beta$ real, the  condition
$\,\,\overline{g_{1\bar{2}}}\equiv 0$ implies
from (\ref{ch}), (\ref{real2})  that $g_{1\bar{2}}=g_{1\bar{1}}/2=1$.

\noindent The $tt^*$ equations  reduce to
\eqn\label{diag}
\del_{\bar{\alpha}}\del_{\alpha}\ln{g_{0\bar{0}}}&=&
-{2\over{|\alpha|^2}}\,{g_{0\bar{0}}}^{-1}+{1\over{2|\delta|^2}}
(|\delta|^2 \pm 1)\,{g_{0\bar{0}}}\no\\
\del_{\bar{\beta}}\del_{\beta}\ln {g_{0\bar{0}}}
&=& \pm 2|\alpha|^2|\delta|^2{g_{0\bar{0}}}\, -\,
{1\over {2|\beta|^2}}\,\left(1 \pm |\delta|^2\right)\,{g_{0\bar{0}}}^{-1}\no\\
\del_{\bar{\beta}}\del_{\alpha}\ln {g_{0\bar{0}}}
&=&2\alpha^*g_{0\bar{0}}-{2\over{\alpha\beta^*}}{g_{0\bar{0}}}^{-1}
\enq
where $|\delta|^2=1/(4\beta -1)$ if $\beta>1/4$ (and the $+$ sign is chosen
in $\pm$), and
$|\delta|^2=1/(1-4\beta)$ if $\beta<1/4$ (and the $-$ sign is chosen).

We now solve for $\beta>1/4$. The solution for $\beta<1/4$ is similar
and the conclusions the same.
\noindent Redefining for the first of (\ref{diag})
\eq\label{eq1}
z=4\alpha^{1/2}\beta^{1/4},\quad
 \ln {g_{0\bar{0}}}= {{q_0}}- \ln |\alpha| +\ln d,
\quad  d=2|\delta|/\sqrt{|\delta|^2+1}=\beta^{-1/2}\no
\en
we get $$\,\, \del_{\bar{z}}\del_{z}{{q_0}}=\sinh {{q_0}}$$
\noindent The regular solution exists for $r=2$.
Then, we can predict the behavior of the metric for fixed $\beta$.
As $z\rightarrow 0$ (or $|\alpha|\rightarrow 0,\,r_1\rightarrow \infty$),
\eqn
g_{0\bar{0}}&\simeq& {d\over{|\alpha|}}{|z|^2}\left
(-\log {{|z|}\over 2}-\gamma\right )^2 f(\beta)|z|^m + \dots\no\\
&\simeq &16\left[-\ln \sqrt{2}|\alpha|^{1/2}|\beta|^{1/4}-\gamma\right]^2
f(\beta)|z|^m\no
\enq
where $f(\beta)$ is some integration function.

For $|\alpha|\rightarrow \infty,\,$ ( $r_1\rightarrow -\infty)$,
\eq
g_{0\bar{0}}\simeq {1\over{|\alpha|\beta^{1/2}}}\left[1-{1\over{\sqrt{\pi}}}
{1\over{|\alpha|^{1/4}\beta^{1/8}}}
\exp{\left(-8\sqrt{|\alpha|}\beta^{1/4}\right)}\right]f(\beta)|z|^m.
\en
The second equation of (\ref{diag})  will give us
 the behavior of the metric near the singularity.
\noindent Redefining
$\qquad \ln {g_{0\bar{0}}}= {{q_0}}-{1\over 2}\ln |\beta|-\ln |\alpha|,$
we can write
\eq\label{eq2}
 {{1}\over{4|\alpha|}}\beta^{1/4}\bar{\beta}^{1/4}(4\beta-1)^{1/2}
(4\bar{\beta}-1)^{1/2}\del_{\bar{\beta}}\del_{\beta}{{q_0}}=\sinh {{q_0}}.
\en
Now we look for a change of variables such that
\eq
{1\over {2\alpha^{1/2}}}\beta^{1/4}(4\beta-1)^{1/2}\del_{\beta}={{d\beta}
\over{dw}}\del_{\beta}\equiv\del_w\no
\en
So we need to find the function
\eq\label{ell1}
w(\beta)=\int{{2\alpha^{1/2}d{\beta}}\over{\beta^{1/4}(4\beta-1)^{1/2}}}\no
\en
Now the integral becomes, under the  substitution $z=\sqrt{4\beta-1}$
\eq
\sqrt{2}\alpha^{1/2}\int {{dz}\over{(1+z^2)^{1/4}}}\no
\en
which, upon setting $z=\sinh x$, transforms to
\eq
\sqrt{2}\alpha^{1/2}\int \sqrt{\cosh x}dx\no
\en
whose solution is in terms of elliptic integrals\cite{grad}
\eq\label{ell}
w=\sqrt{2}\alpha^{1/2}\left\{{\sqrt{2}}\left [F(\kappa,
1/\sqrt{2})-2E(\kappa,1/\sqrt{2})\right] +
{{2\sinh x}\over{\sqrt{\cosh x}}}\right\},
\en
where $F(\kappa,r)$ and $E(\kappa,r)$ are elliptic integrals of the first
and second kind and
\eq
\kappa = \arcsin\sqrt{{\cosh x -1}\over{\cosh x}}.\no
\en
We see  that when $\beta\rightarrow 1/4$,
$\kappa\rightarrow {x\over{\sqrt{2}}}\sim{1\over{\sqrt{2}}}(4\beta-1)^{1/2},
\quad F(\kappa)$ and $E(\kappa)\sim \kappa,$ and
$w\sim \sqrt{2}\,\alpha^{1/2}(4\beta - 1)^{1/2}$.

\noindent Now, as $w\rightarrow 0$, the solution of (\ref{eq2}) for the metric
is
\eqn
g_{0\bar{0}}(|w|)&\simeq & {1\over{|\alpha||\beta|^{1/2}}}|w|^2
\left[-\left (\log{|w|\over
2}+\gamma\right )\right]^2\,h(|\alpha|)|w|^m +\dots \no
\enq
and for $|w|\rightarrow \infty$,
\eqn\label{inf1}
g_{0\bar{0}}(|w|)&\sim & {1\over{|\alpha||\beta|^{1/2}}}\left[
1-{2\over{\sqrt{\pi|w|}}}\exp[-2|w|]\right]q(|\alpha|)|w|^p
\enq
with $h(|\alpha|),\,q(|\alpha|)$ some integration functions.
We notice here, with $|\alpha|$ large,
the appearance of solitons of  mass $m=2|w|\sim
2\sqrt{2}|\alpha|^{1/2}(4\beta -1)^{1/2}$ which become massless
as $\beta$ approaches the singular point in the moduli,
$\beta\rightarrow 1/4$.

Altough we haven't obtained the full solution of the $tt^*$ system,
the behavior of the metric near the singularity is now clear: for $\beta$
  close to $1/4,\,$ the function
$|w|\sim \sqrt{2}|\alpha|^{1/2} |1-4\beta|^{1/2}$
and  the metric behaves as
$g\sim |w|^2[-(\,\log |w|const.\, +\,\gamma)]^2.\,$
 Near $|\alpha|\rightarrow \infty$, we flow to a conformal model with massless
solitons.

\sect{Conclusions}

We have studied supersymmetric $\sigma$-models on a particularly simple
example of a toric variety. However, this example allowed us  to explore
the physics around the singular locus of the model.
 By using the description of these models
as  gauged linear $N=2$ $\sigma$-models and
 the methods of topological-antitopological fusion, we have shown that
the model contains solitons which become massless at the singular point
where one of the gauge symmetries is unbroken.
 At that point,  the model consists of both a Higgs phase
with massless chiral fields and a Coulomb phase.\cite{mopl}
We have here an example of what can happen in $N=2$ models in 2D as
described in (\cite[x,y,z]{cv2,cfiv,vafa1}), i.e. the jumping in the
number of Bogomolnyi solitons.
 These solitons are related to the intersection
numbers of the vanishing cycles of the singularity. These intersection
numbers undergo a jump and the number of BPS solitons changes as
a result of the monodromy of the vanishing cycles.
This is also the suggestion of what happens for models
in $D=4$.\cite{vafa1,seiwit,stro}
We can also view our example as an example of flows in massive theories.
We note that we are flowing in the infra-red limit to another conformal
theory with massless solitons and it would be interesting to determine
this CFT. We leave this as an open problem for future work.

We can hope that what we
have learned in this simple case can be generalized to explain more
sophisticated models.
In \cite[x]{mopl}, the authors go on to explore the
superconformal nonlinear sigma models with Calabi-Yau target spaces that can
be embedded as hypersurfaces  in toric varieties.
They find there again that  the singularities divide up
the parameter space in different phases. The present work then also
suggests that if one encounters the same types of singularities as the one we
have studied, a similar explanation in terms of massless
 solitons appearing at the boundaries of the
different phases  of the model is plausible, as has been discussed
recently in the litterature.\cite{cand,cggk,gms}

\sect{Acknowledgments}

I would like to thanks P. Aspinwall and A. Klemm for explaining
me about toric varieties and M. Douglas for useful comments.
This work was supported by a Cern Fellowship and
a Oxford University Glasstone fellowship.
\goodbreak

\end{document}